\documentclass[%
 preprint,
 amsmath,amssymb,
 aps, physrev,
 superscriptaddress
]{revtex4-2}

\usepackage{graphicx}
\usepackage{dcolumn}
\usepackage{bm}
\usepackage{hyperref}
\usepackage{siunitx}  
\usepackage{gensymb}
\usepackage{soul}
\usepackage{pifont}
\usepackage{enumitem}
\usepackage{natbib}
\usepackage{longtable}
\usepackage{array}

\usepackage{xcolor}  

\usepackage{wrapfig}  
\usepackage[mathlines]{lineno}

\usepackage{graphicx}
\graphicspath{{Figures/}}

\begin{document}

\preprint{APS/123-QED}

\title{\textbf{Vision AI Agent for Continuous Material Monitoring of LEGEND-1000 LoFi Reentrant Tube} 
}%

\author{Sonata Simonaitis-Boyd}
\email{sonata@ucsd.edu}
\affiliation{Halıcıoğlu Data Science Institute, University of California San Diego, 9500 Gilman Dr., La Jolla, CA 92093, USA}
\author{Soonhong Lee}
\affiliation{Los Alamos National Laboratory, Physics Division, PO Box 1663, MS: H803 Los Alamos, NM 87545, USA}
\author{Lauren N. O'Brien}
\affiliation{Los Alamos National Laboratory, Physics Division, PO Box 1663, MS: H803 Los Alamos, NM 87545, USA}
\author{Brandon T. Turner}
\affiliation{Los Alamos National Laboratory, Physics Division, PO Box 1663, MS: H803 Los Alamos, NM 87545, USA}
\author{Ralph Massarczyk}
\affiliation{Los Alamos National Laboratory, Physics Division, PO Box 1663, MS: H803 Los Alamos, NM 87545, USA}
\author{Steven R. Elliott}
\affiliation{Los Alamos National Laboratory, Physics Division, PO Box 1663, MS: H803 Los Alamos, NM 87545, USA}

\author{Aobo Li}
\affiliation{Halıcıoğlu Data Science Institute, University of California San Diego, 9500 Gilman Dr., La Jolla, CA 92093, USA}
\affiliation{Department of Physics, University of California San Diego, 9500 Gilman Dr., La Jolla, CA 92093, USA}
\author{Alexander F. Leder}
\email{aleder@lanl.gov}
\affiliation{Los Alamos National Laboratory, Physics Division, PO Box 1663, MS: H803 Los Alamos, NM 87545, USA}



\date{\today}

\begin{abstract}
We report on a vision AI agent pipeline for non-contact material strain and property extraction from video data, demonstrated on video taken during hydrostatic testing of four OFHC copper cylinders conducted as part of the LEGEND-1000 hardware validation campaign. Traditional strain gauge measurements proved unreliable, motivating a fully-automated agentic alternative. The agent was built on the LangChain framework with Claude Haiku 4.5 as its central reasoning engine, integrating a specialized suite of computer vision tools: FFmpeg for video preprocessing and rotation correction via Hough Line Transform, the Segment Anything Model 2 (SAM2) for spatiotemporal segmentation with automated memory-informed dynamic chunking, and a hybrid EasyOCR and LLM-based timestamp validation pipeline. Three specialized sub-agents were developed to process the video data and obtain cylinder diameters and timestamps while autonomously handling obstacles such as corrupted frames and memory limits. From the diameter profiles synchronized to pressure data, hoop stress--strain curves were reconstructed and yield strengths were calculated using the 0.2\% offset, 0.5\% EUL, and Johnson-Cook methods across two independent tests. Cross-validation against a non-agentic pipeline confirmed agreement for the diameter extraction at the $\pm$5 pixel level. The material properties and testing results were further compared to Ansys mechanical simulations performed as part of the LEGEND-1000 reentrant tube design campaign. This work showcases the power of agentic pipelines to extract materials data from video alone.
\end{abstract}

\maketitle


\section{Introduction}\label{sec:intro}
Neutrinoless double beta ($0\nu\beta\beta$) decay~\cite{0vbb_review,Dolinski2019} is a lepton-number-violating process ($\Delta L=2$) that would help answer several important outstanding questions in the field of neutrino physics and cosmology. The observation of $0\nu\beta\beta$ decay would prove the Majorana nature of the neutrino and provide an avenue for leptogenesis~\cite{leptogenesis}, a model explaining the observed matter--antimatter asymmetry of the universe. $0\nu\beta\beta$ decay has yet to be observed and presents a difficult experimental challenge, as current limits have placed the lower limit on the half-life at $> 10^{26}$\,yr~\cite{Adams2022,MJDPRL,klz_newresult,GERDAresult,LEGENDPRL-2026}, making $0\nu\beta\beta$ decay, if detected, one of the slowest nuclear decay processes in the universe with the corresponding $2\nu\beta\beta$ decay being the slowest observed process.

The currently-operating Large Enriched Germanium Experiment for Neutrinoless $\beta\beta$ Decay (LEGEND)~\cite{LEGENDPRL-2026} and its precursor experiments---\textsc{Majorana Demonstrator}~\cite{MJDPRL} and GERDA~\cite{GERDAresult}---have placed limits on this process by utilizing a source-detector configuration of high-purity germanium (HPGe) detectors highly enriched in $^{76}$Ge, taking advantage of their proven high energy resolution and low background rates. Next-generation germanium-based experiments, such as LEGEND-1000~\cite{legend_pcdr}, aim to probe half-lives beyond $10^{27}$~years, but achieving this sensitivity requires a tonne-scale array of enriched HPGe detectors and a near-total suppression of background events in the region of interest as discussed in the LEGEND-1000 proposal~\cite{legend_pcdr}. Consequently, the mechanical design of LEGEND-1000 must adhere to strict radiopurity standards while maintaining operational robustness over a decade-long data-taking campaign.

For LEGEND, a unique component is the Reentrant Tube (RT), which is designed to create two separate chambers of Liquid Argon (LAr): an inner most volume of radiopure underground LAr surrounding the HPGe detectors, and an outer volume filled with LAr from atmospheric sources. The RT represents a critical structural component of the LEGEND-1000 reference design: it must operate in a complex cryogenic environment over years without failure or leaks while adhering to the strict radiopurity requirements. To validate the mechanical design, the LEGEND-1000 collaboration plans a series of material tests, starting with the so-called ``LoFi tests'' (``LoFi'' standing for Low Fidelity), to quantify the RT material strain, determine the optimal e-beam welding parameters, and test manufacturing techniques. For the first of these tests, we sought to measure the strain of the test cylinders until the weld seams failed by attaching strain gauges to the surface of the cylinders. However, the telemetry data collected during these tests proved to be incomplete and constricted to areas directly adjacent to the strain gauges. This resulted in the need for a non-contact method to extract the needed material and weld parameters across a wider range of the cylinder body.

To address this challenge, we introduce a novel analysis approach: a Vision AI Agent designed for continuous material strain estimation. In traditional experimental data analysis, software pipelines rely on static C++ or Python scripts; they execute a predetermined sequence of commands that often struggle to adapt to unexpected variations in the data. An AI agent overcomes this limitation by acting as an autonomous system capable of perceiving its environment, reasoning about its current state, and executing targeted actions to achieve specific objectives~\cite{Yaoetal2022}. The agent iteratively observes incoming video data, decomposes complex analysis tasks into manageable sub-steps, and autonomously selects the most appropriate algorithms to complete each step.

In this work, we developed an agent that uses a Large Language Model (LLM) as its central reasoning engine. Rather than relying on a rigid processing chain, the LLM acts as an autonomous controller that dynamically selects and uses specific computer vision tools to analyze the LEGEND LoFi test videos~\cite{Shenetal2023}. Guided by natural language instructions (prompts), this agentic workflow establishes a highly-flexible, fully-automated analysis pipeline capable of adapting to the nuances of experimental data.

This paper is structured as follows: Section~\ref{sec:LoFi-test} details the experimental setup of the RT LoFi testing and the specific failure mode of the conventional sensors. Section~\ref{sec:AI-workflow} describes the architecture of the Vision AI Agent developed to retrieve the tube geometry, with the results and comparison to Ansys mechanical simulations presented in Section~\ref{sec:agentic_result}. Section~\ref{sec:improvements} outlines improvements for any future measurement runs. Lastly, Section~\ref{sec:conclusion} describes the conclusion and outlook. This work represents one of the first applications of an agentic Vision AI workflow to automate critical hardware validation in LEGEND and cross-check mechanical simulations of the RT design, establishing a precedent for autonomous optical monitoring in rare-event searches where traditional instrumentation may prove difficult to implement or have a high failure rate.

\section{The LEGEND 1000 Experiment - LoFi Testing Campaign}\label{sec:LoFi-test}
The LEGEND-1000 experiment represents a large jump in scale over previous HPGe $0\nu\beta\beta$ decay searches~\cite{GERDAresult,MJDPRL}. Part of this scaling involves the commissioning of a new large-scale component, the RT, in order to meet the science goals of the LEGEND-1000 program~\cite{legend_pcdr}. The RT is a 6-meter-tall, 2-meter-diameter, and 1.5-millimeter-thick cylinder (at its thinnest part), designed to create an inner designated ultra-clean LAr volume separate from the larger, less-radiopure outer LAr volume. The RT itself must also satisfy stringent requirements for radiopurity, mechanical stability, and longevity while operating inside of this LAr cryogenic environment at 87 K. As part of the design phase, the LEGEND-1000 experiment has created a testing campaign using a series of progressively larger test stands to specifically test the viability of each of the stringent requirements of the experiment. For the first of these tests, we manufactured a total of six small-scale ``LoFi test'' cylinders using commercially-available high-purity oxygen-free high-conductivity (OFHC) copper e-beam-welded by pro-beam GmbH~\footnote{\href{https://www.pro-beam.com/}{pro-beam GmbH \& Co.\ KGaA}, Lindenallee 22, 39288 Burg, Germany.}. These LoFi cylinders were specifically designed to test the e-beam welding parameters on OFHC copper and benchmark the Ansys~\cite{ansysmech2024} mechanical stress--strain parameters simulated as part of the overall RT design process. 

The primary goal of the LoFi testing campaign was to experimentally characterize the following components of the material/manufacturing process for OFHC copper vital for confirming the overall RT design:
     \begin{enumerate}
         \item \textbf{Weld Optimization}: Identify a set of e-beam welding parameters that consistently produce strong, clean, and reproducible joints in OFHC copper that meet both mechanical and radiopurity constraints.
         \item \textbf{Mechanical Characterization}: Determine the structural limits of the OFHC copper material and the e-beam welding seams. This includes measuring the deformation behavior under pressure and identifying the conditions under which the welds and/or material fail. At the same time, these tests were designed to test the reproducibility of the above extracted material/weld parameters and refine material simulations performed as part of the RT design campaign.
         \item \textbf{Simulation Verification}: As part of the LEGEND RT design process, a series of structural simulations were run to determine the overall viability of any given design. Material property data on specialty materials, such as OFHC or underground electroformed copper, are rare and need to be tested before the parameters are input into any mechanical simulation. 
     \end{enumerate}

\begin{figure}[htbp]
    \centering
    \begin{minipage}[b]{0.22\textwidth}
        \centering
        \includegraphics[height=6cm]{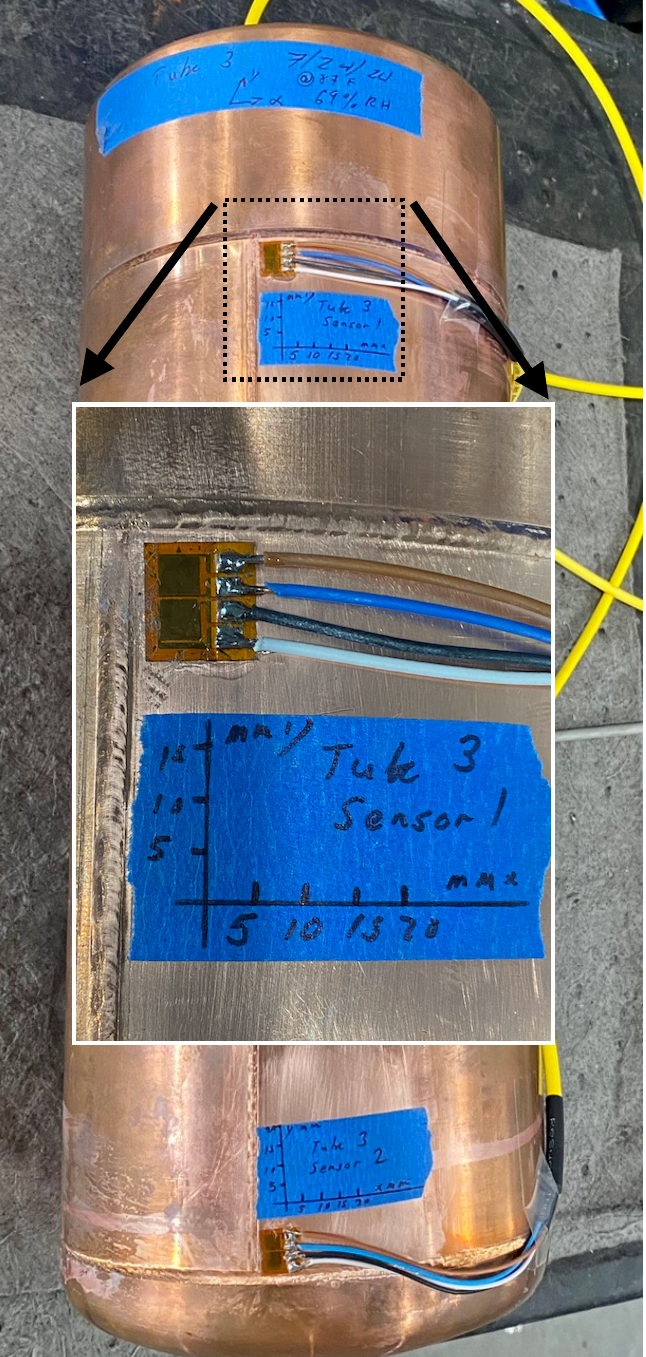}
        \\ (a)
    \end{minipage}
    \hfill
    \begin{minipage}[b]{0.75\textwidth}
        \centering
        \includegraphics[height=6cm]{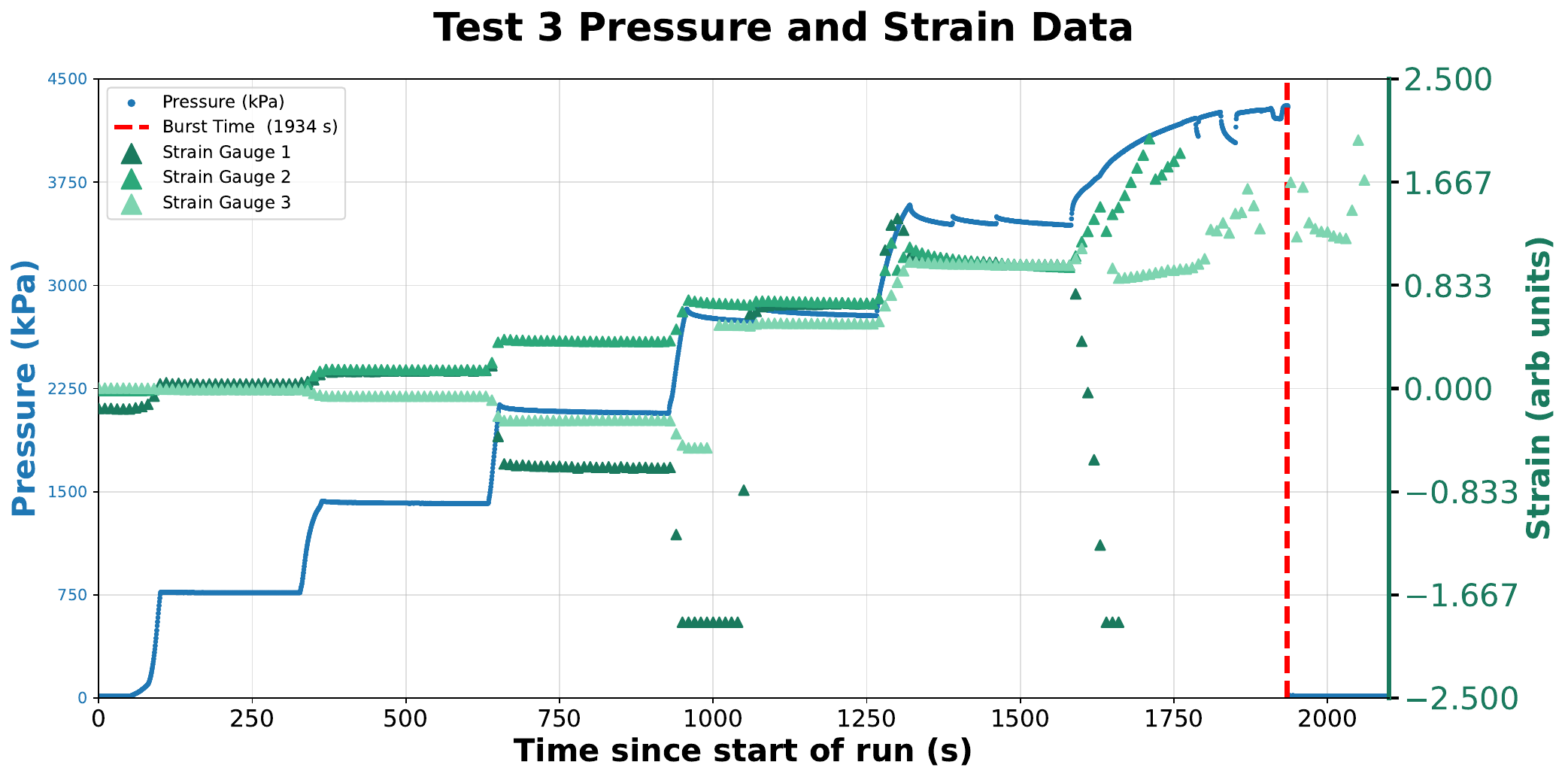}
        \\ (b)
    \end{minipage}
    \caption{(a) Example positioning of the strain gauge sensors placed near 
 the intersection of the weld seams (inset). For each test there were three 
 sensors placed around each test cylinder, including one around the back (not 
 pictured). (b) Raw strain gauge data from Sensors 1--3 during Test 3 along 
 with associated pressure data. While the strain data has periods where it 
 follows the pressure curve, there are numerous data artifacts such as sudden 
 drops, lack of data, and even negative strain data values that are visible.}
    \label{fig:sensordata}
\end{figure}

These tests were performed at the Houston, TX facility of Pneumatic and Hydraulic Company, LLC~\footnote{\href{https://pneumaticandhydraulic.com/}{Pneumatic and Hydraulic Company, LLC Website}, 20500 Clay Center Dr, Katy, TX 77449, USA.} who specialize in the design, construction, installation, servicing, and testing of high-pressure systems for various industries including oil and gas, aerospace, military, petrochemical, and marine. One of the first metrics derived from these tests was the burst pressure---the internal pressure at which the LoFi cylinders ruptured during hydrostatic testing. A total of four cylinders were tested during this campaign, with each test labeled sequentially (Tests 1--4), between September and November 2024 and testing taking place in a hardened steel containment vessel. The combined burst pressures across all four tests were recorded as $4286 \pm 11$~kPa, with the small error indicating very reproducible burst pressures across all tests. For the tests that were video recorded, data were logged with multiple cameras positioned at different angles to visually document each test. The initial pressure ramping was to occur step-wise over several hours to allow for any internal material stresses to dissipate and equilibrate (see Fig.~\ref{fig:sensordata}b). Each LoFi cylinder was also equipped with three 2D strain gauges (supplied by Micro-Measurement~\footnote{\href{https://micro-measurements.com/}{Micro-Measurements}, 951 Wendell Blvd., Wendell, NC 27591, USA.} and seen in Fig.~\ref{fig:sensordata}a) to monitor real-time deformation and measure strain along two orthogonal axes with the aim of capturing both longitudinal and transverse deformation during testing. 


\vspace{-2.5em}
 
\subsection{Initial Results}\label{sec:LoFi-initresults}
In each test, the burst moment was clearly identifiable in both the pressure logs and the camera footage and was used as a synchronization point across all subsequent analyses. The strain gauge data however did not show any clear or consistent correlation to the internal pressure (see Fig.~\ref{fig:sensordata}). In particular, the strain gauge data showed numerous artifacts that were inconsistent with the pressure data trends. For example, there were time periods where the strain gauge displayed rapid instability in the data during pressure ramping or where the gauge recorded a negative value. Additionally, certain strain gauge sensors detached from the LoFi cylinder during testing, resulting in no data being recorded. There were time periods where the strain gauge did track fairly closely with the pressure data (see Gauge 2 in Fig.~\ref{fig:sensordata}), however the specific sensor and exact time range where the data matched the pressure trends varied widely from test to test. Test 1 only had one working sensor to record data while Test 3 had data from all sensors, though these each showed different trends in the strain curve. Fortunately, video recordings of the LoFi RT during the first and second tests were available, which offered an alternative path forward to provide material data to develop and validate a novel, non-contact method for strain extraction.


The core idea was to use frame-by-frame analysis of the video footage to track the real-time deformation of the cylinder's outer diameter under increasing internal pressure. By correlating these time-resolved diameter measurements with the corresponding pressure logs, it became possible to reconstruct the mechanical strain profile of the vessel as a function of time. Discussion of the cylinder's strain--stress parameters is continued in Section~\ref{sec:agentic_result}. 





\section{Vision AI Agent}\label{sec:AI-workflow}

The Vision AI agent presented in this work was implemented using the LangChain~\cite{Chase_LangChain_2022} framework. We utilized the Claude Haiku 4.5 Large Language Model (LLM)~\cite{anthropic2024claude3} as the central reasoning engine, pairing it with a specialized suite of computer vision tools including video processing, Optical Character Recognition (OCR)~\cite{JaidedAI_EasyOCR_2020}, and the Segment Anything Model 2 (SAM2)~\cite{ravi2024sam2}. The LLM used these tools to extract the LoFi cylinder's diameters as a function of time, effectively transforming raw visual feeds into precise time-series measurements. In this section, we first detail the individual computer vision tools employed, followed by a discussion of the agent's architectural construction and workflow logic. While terms will be defined as necessary in this text, a glossary of relevant AI-related terms can be found in Appendix~\ref{app:glossary}.

\subsection{Tool Preparation}\label{sec:tools}
The Vision AI Agent developed in this work integrates the SAM2 vision foundation model with external dependencies. SAM2 extends the transformer-based architecture of its predecessor to the temporal domain, enabling unified video and image segmentation. Unlike simple frame-by-frame approaches, SAM2 utilizes a streaming memory mechanism to propagate segmentation masks across video frames, ensuring spatiotemporal consistency for dynamic objects. Complementing this foundation model, we leveraged high-performance multimedia framework FFmpeg~\cite{FFmpeg} to handle the efficient decoding and preprocessing of raw high-resolution video files. Finally, we used EasyOCR~\cite{JaidedAI_EasyOCR_2020} for first-stage optical character recognition. OCR is defined as the computational process of converting text embedded within visual media into machine-readable string formats. In this pipeline, OCR was critical for extracting the timestamp from each video frame, enabling the agent to synchronize the tube diameters with the video timeline.

We developed tools built upon these dependencies, which the LLM is tasked with invoking within the agentic workflow. By interpreting provided natural language descriptions, the model determines the appropriate context and method for execution. A full list of pipeline tools and their corresponding natural language descriptions is shown in Appendix~\ref{app:tool_list}.

\subsection{Agent Design}\label{sec:AgentDesign}
\begin{figure}
    \centering
    \includegraphics[width=\textwidth]{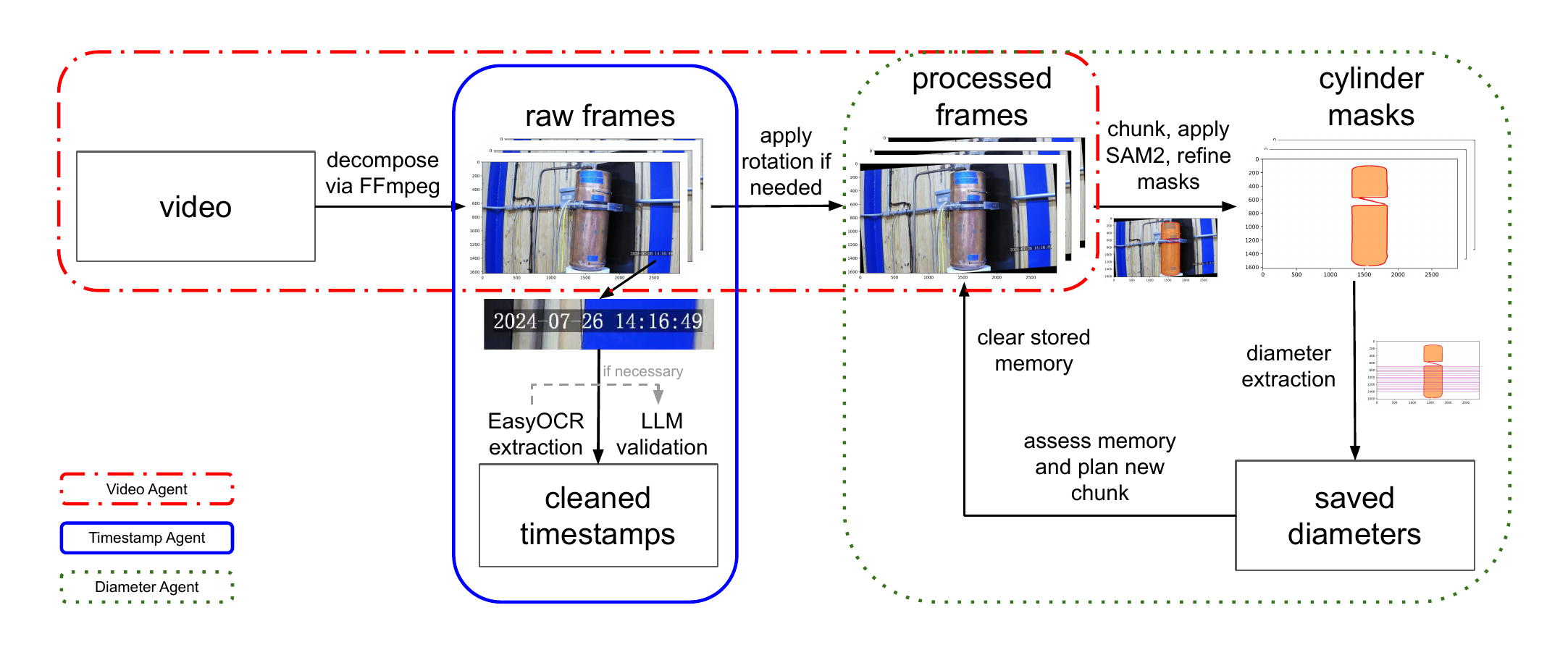}
    \caption{Diagram of the Vision AI Agent, a workflow composed of three AI subagents: the Video Agent (red), Timestamp Agent (blue), and Diameter Agent (green).}
    \label{fig:agent-pipeline}
\end{figure}

The full agentic pipeline presented in this paper consists of three subagents: (i) the Video Agent, (ii) the Diameter Agent, (iii) the Timestamp Agent. These subagents operated in a stage-wise synergistic manner to carry out the complete video analysis tasks. An overview of the full agentic pipeline is presented in Fig.~\ref{fig:agent-pipeline}.

Each AI agent was designed to execute a specific set of tasks by using the tools described in Section~\ref{sec:tools} and Appendix~\ref{app:tool_list}. The operational logic of each agent is defined by a system prompt---a set of natural language instructions that guide the agent on task objectives and appropriate tool selection. Acting as the central reasoning engine, the LLM interprets this system prompt to autonomously interface with the toolset and orchestrate the workflow.

\subsubsection{Video Agent}
The first component of our workflow is the Video Agent, which was tasked with the initial preprocessing of raw LoFi testing footage. This agent operates as a highly-configurable pipeline, allowing the LLM to define parameters such as frame extraction rates, output directory structures, and visualization. To execute these tasks, the agent is equipped with a specific set of tools: \texttt{decompose\_video\_to\_frames}, which utilizes FFmpeg for efficient decomposition; \texttt{calculate\_rotation\_angle}, which computes the necessary angular adjustment to vertically align the target object (in this case the LoFi cylinder); and \texttt{rotate\_all\_frames}, which applies this correction to the dataset.
A critical function of the Video Agent is rotation correction of video frames. The ``Test 1'' dataset was filmed at an oblique angle requiring correction to restore the tube's upright geometry. The Video Agent addressed this by employing a Hough Line Transform~\cite{Duda1972} to detect structural edges; these edges were then used to calculate the precise rotation angle required to straighten the tube, after which the transformation was applied globally across the dataset.




\subsubsection{Diameter Agent}
The Diameter Agent is the most sophisticated component of the workflow, tasked with supervising the core segmentation and analysis pipeline. It operates in the following steps: (i) initialization of the SAM2 inference state, (ii) mask refinement, (iii) temporal propagation, and (iv) final diameter extraction. Following the output of the Video Agent, the Diameter Agent begins its tasks by refining the segmentation mask through the provision of positive and negative ``clicks'' which guide the model to include or exclude specific regions of the frame. Once the initial mask is defined, it is propagated across the entire set of frames for the chunk. From these segmented masks, the agent generates an outline and determines diameters by calculating the pixel distance between the left and right mask edges at specific user-defined vertical positions (z-levels).

\begin{figure}[htbp]
    \centering
    \begin{minipage}[b]{0.24\textwidth}
        \centering
        \includegraphics[height=5cm]{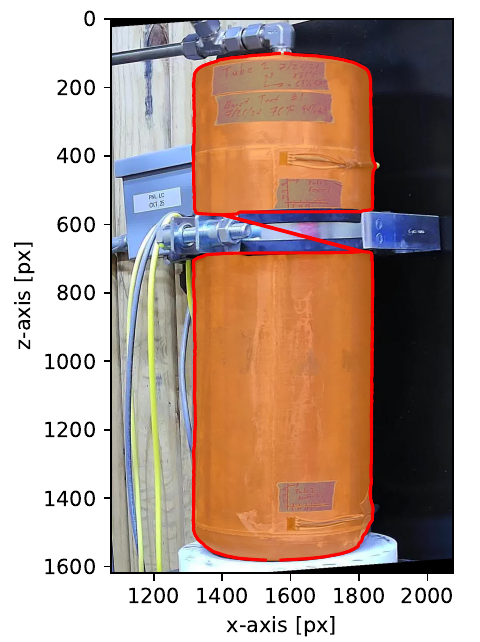}
        \\ (a)
    \end{minipage}
    \hfill
    \begin{minipage}[b]{0.24\textwidth}
        \centering
        \includegraphics[height=5cm]{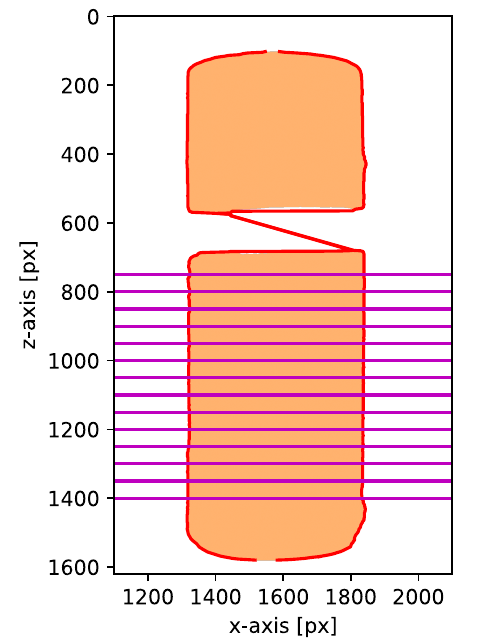}
        \\ (b)
    \end{minipage}
    \hfill
    \begin{minipage}[b]{0.24\textwidth}
        \centering
        \includegraphics[height=5cm]{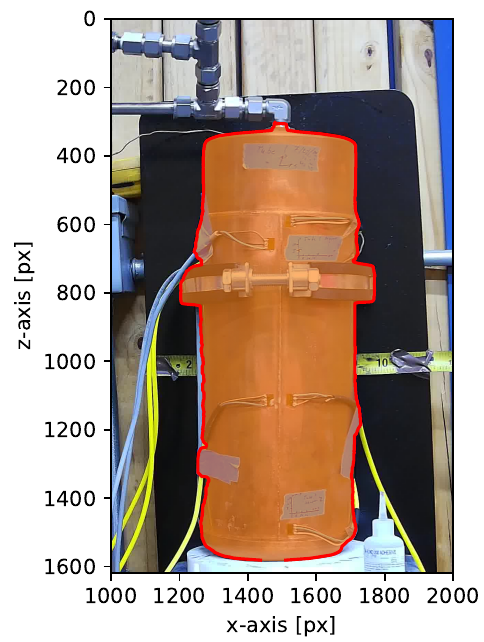}
        \\ (c)
    \end{minipage}
    \hfill
    \begin{minipage}[b]{0.24\textwidth}
        \centering
        \includegraphics[height=5cm]{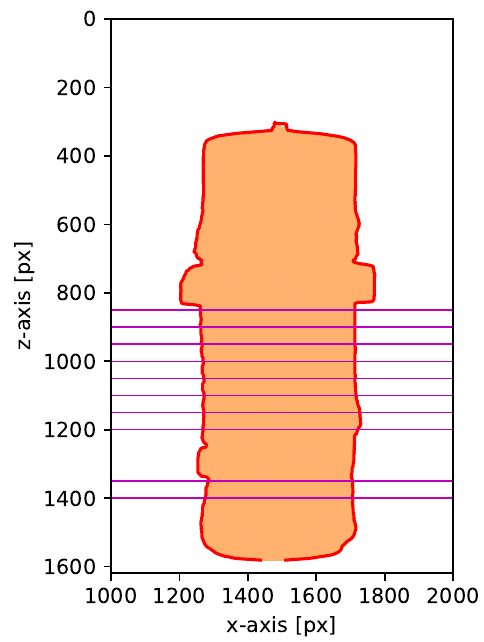}
        \\ (d)
    \end{minipage}
    \caption{(a) A raw frame from the Test 1 video, overlaid with the refined 
segmentation mask (orange) and its outline (red). (b) The mask and outline 
overlaid with horizontal lines (magenta) corresponding to z-levels 700--1400 
in increments of 50. (c) Test 2 raw video frame overlaid with the refined 
segmentation mask (orange) and its outline (red). (d) The mask and outline 
overlaid with horizontal lines (magenta) corresponding to z-levels 850--1200 
and 1350--1400 in increments of 50.}
    \label{fig:test1outlines}
\end{figure}

The Diameter Agent was applied to both the Test 1 and Test 2 datasets. In the Test 1 analysis shown in Fig.~\ref{fig:test1outlines}ab, diameters were measured at pixel z-levels from 750 to 1400 with an increment of 50 pixels. This approach allowed us to study tube deformation at various heights. For the Test 2 analysis shown in Fig.~\ref{fig:test1outlines}cd, we specified z-levels of 850--1200 and 1350--1400 in increments of 50 pixels. Notably, the rotational angle correction was omitted for the ``Test 2'' dataset, as the initial camera alignment was determined to be sufficiently upright for accurate diameter extraction.

A key challenge in running the Diameter Agent over ultra-long video datasets is the prevalence of CUDA Out Of Memory (OOM) errors. In non-agentic iterations of this pipeline, addressing this obstacle required manual ``chunking'' of frames and frequent kernel restarts to forcibly clear GPU memory. To overcome this agentically, the Diameter Agent implements an automated memory-informed dynamic chunking strategy: the agent groups $n$ frames into a chunk and processes them together, while actively monitoring peak memory usage during intensive steps like mask propagation. Based on this information, the agent dynamically adjusts the size of subsequent chunks based on memory capacity---if memory headroom is high, $n$ is increased to improve processing speed; if the memory exceeds a preset threshold, the agent immediately halts the process and clears all memory. Chunk size $n$ is then reduced and the agent autonomously retries the processing step. After the successful processing of each chunk, the agent calls memory management tools to clear caches and delete redundant variables, freeing up space for subsequent processing.

\subsubsection{Timestamp Agent}
The final component of our pipeline, the Timestamp Agent, was tasked with synchronizing the extracted tube diameter profiles with their corresponding time recorded by the camera. This required the precise retrieval of alphanumeric timestamps embedded in the bottom-right corner of each video frame (e.g., see Fig.~\ref{fig:agent-pipeline}). While standard Python-based OCR tools such as EasyOCR performed adequately for the high-contrast Test 1 dataset, they struggled significantly with the Test 2 video as the poor timestamp--background contrast led to frequent alphanumeric misinterpretations (e.g., misidentifying the number ``9'' as the lowercase letter ``g''). Conventional regular expression (Regex) cleaning can correct formatting issues but cannot resolve these fundamental character misclassifications, necessitating a more robust vision-based approach.

To address this, the Timestamp Agent employs a hybrid multi-stage pipeline designed to balance accuracy with token cost. The workflow begins by cropping out the specific region of interest that contains the timestamp information via the \textsc{set\_timestamp\_region} tool. The agent then executes \textsc{extract\_timestamps\_with\_easyocr}, which retrieves the raw text and processes it via regular expressions to standardize formatting and merge fragmented strings. The output is validated against the expected timestamp format (YYYY-MM-DD HH:MM:SS); if the string contains invalid characters or structural anomalies that Regex cannot resolve, the frame is flagged as problematic. The key innovation of the Timestamp Agent is its hierarchical validation mechanism. Since utilizing an LLM to process and analyze every video frame would be prohibitively expensive, the Timestamp Agent reserves this resource solely for the flagged frames. When a frame is flagged as problematic, the agent invokes the \textsc{validate\_timestamp\_with\_llm} tool, which feeds the image frame directly to the LLM's vision encoder to extract the correct timestamp with high fidelity. 

\section{Agentic Analysis Result}\label{sec:agentic_result}

\begin{figure}[htb!]
    \centering
    \begin{minipage}[b]{0.48\textwidth}
        \centering
        \includegraphics[width=\textwidth]{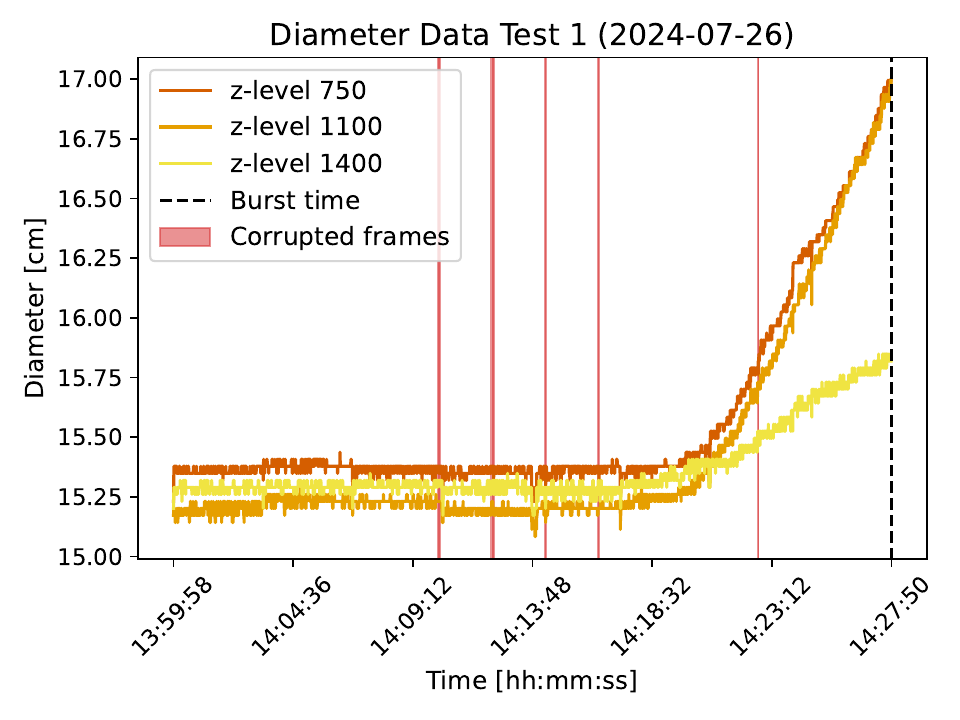}
        \\ (a)
    \end{minipage}
    \hfill
    \begin{minipage}[b]{0.48\textwidth}
        \centering
        \includegraphics[width=\textwidth]{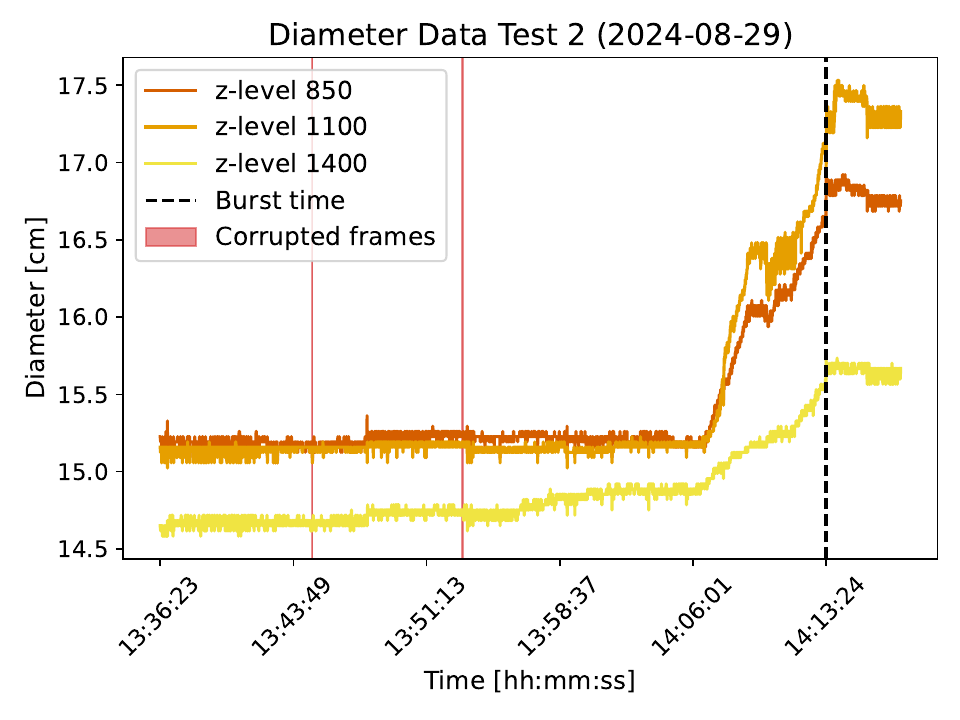}
        \\ (b)
    \end{minipage}
    \hfill
    \begin{minipage}[b]{0.48\textwidth}
        \centering
        \includegraphics[width=\textwidth]{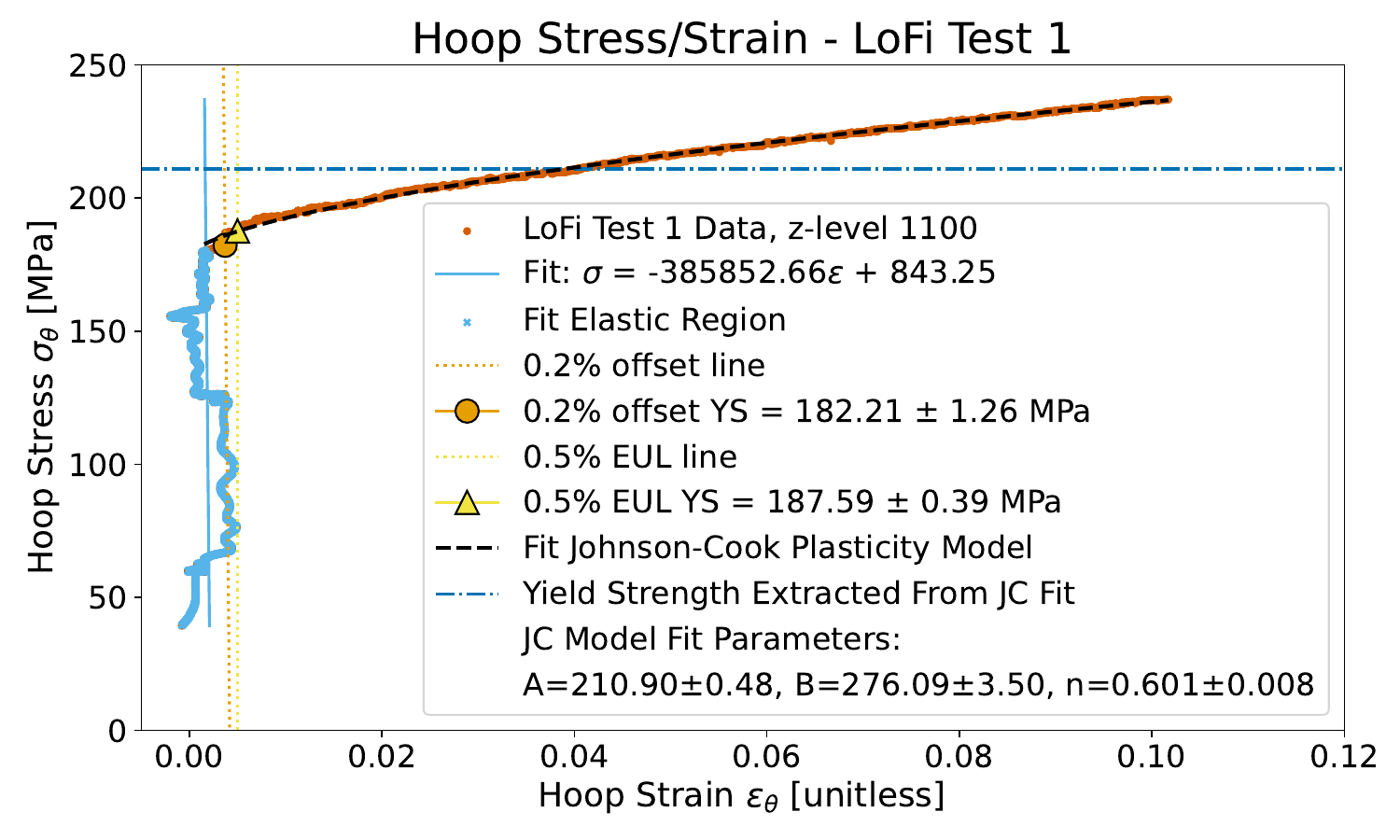}
        \\ (c)
    \end{minipage}
    \hfill
    \begin{minipage}[b]{0.48\textwidth}
        \centering
        \includegraphics[width=\textwidth]{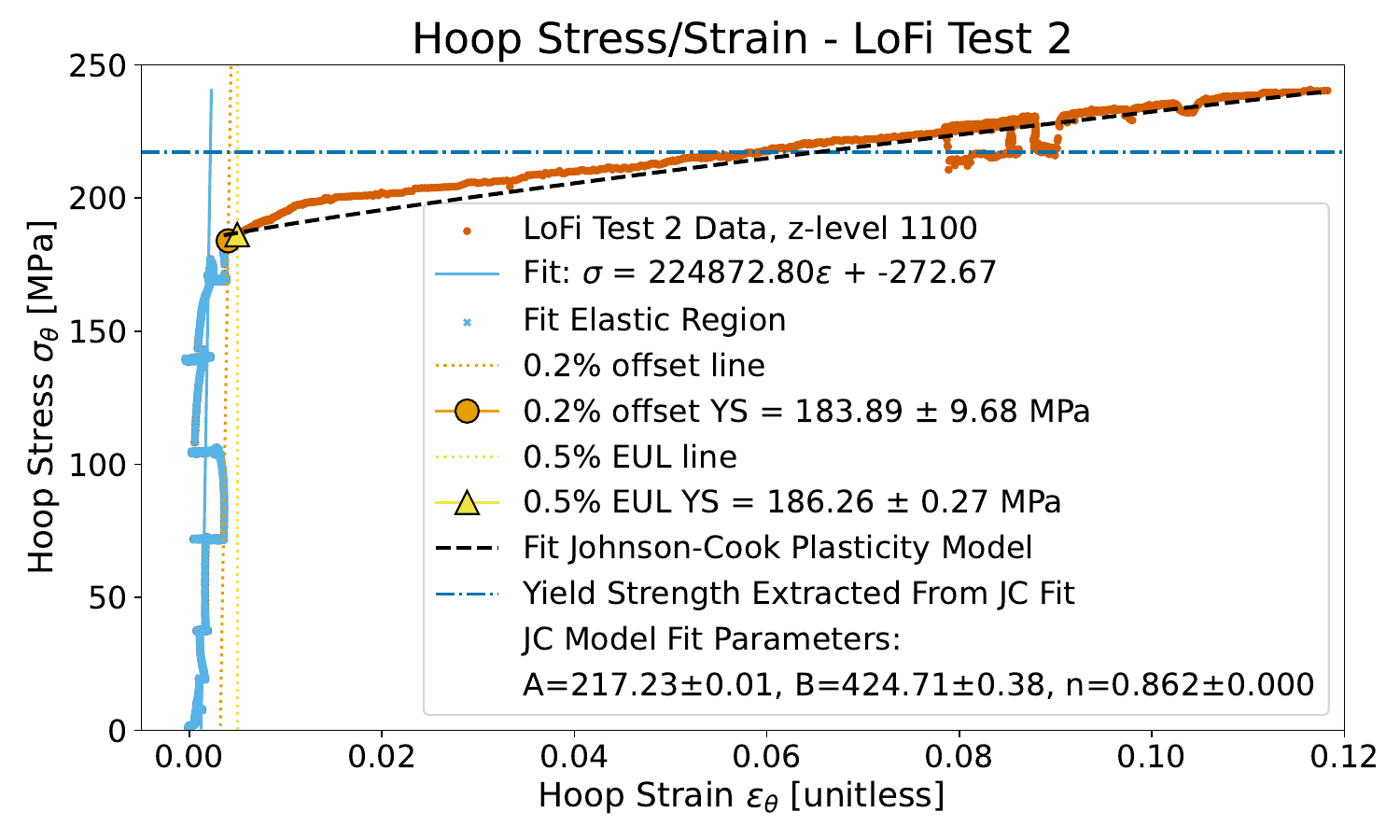}
        \\ (d)
    \end{minipage}
    \caption{(a), (b) Time evolution of the LoFi cylinder centimeter diameters 
for Test 1 and Test 2. In the top panels, diameters are tracked at three 
representative heights: the ``top'' region near the clamp (red), the middle 
region (orange), and the ``bottom'' region near the bottom (yellow). The 
specific pixel z-levels correspond to 750, 1100, and 1400 for Test 1, and 850, 
1100, and 1400 for Test 2. The vertical dashed black line denotes the moment of 
cylinder burst, after which extraction was unreliable. The red masked regions 
correspond to sections of corrupted frames. (c), (d) Representative hoop stress--strain curves 
of the LoFi cylinders for Test 1 and Test 2 with the Johnson-Cook plasticity 
model~\cite{Stopka, Voyiadjis2005} overlaid with parameters fit to the data. The yield strength is measured in three ways: $0.2$\% offset (engineering standard), $0.5$\% EUL, and the $A$ parameter from the Johnson-Cook model fit.}
    \label{fig:diameters_result}
\end{figure}

By synthesizing the timestamps from the Timestamp Agent with the diameters from the Diameter Agent at various z-levels, we obtained the final result of the agentic analysis, as visualized in Fig.~\ref{fig:diameters_result}. The results align with the expected behavior: under continuously-increasing pressure, the test cylinder undergoes significant expansion eventually leading to a burst. The figure captures the start of deformation at approximately 14:18 in Test 1 and 14:06 in Test 2.

The multi-z-level extraction scheme enabled a granular observation of the test cylinder's deformation topology: the diameter increased most rapidly at the middle z-levels---corresponding to the ``belly'' of the LoFi cylinder---while the change near the top and bottom remained comparatively slow. 

\begin{table}[ht!]
    \centering
    \begin{ruledtabular}
    \begin{tabular}{lllllll}
        \textbf{Z-level} & \textbf{0.2\% YS} & \textbf{0.5\% YS} &  \textbf{JC YS} & \textbf{Pixel error} & \textbf{$\epsilon_{\theta}$ error} & \textbf{$\sigma_{\theta}$ error} \\
         & \textbf{[MPa]} & \textbf{[MPa]} & \textbf{[MPa]} & \textbf{[cm]} &  & \textbf{[MPa]} \\
        750 & $190.75 \pm 2.27$ & $194.83 \pm 0.24$ & $216.38 \pm 0.01$ & 0.019 & 0.0085 & 0.2354 \\
        1100 & $182.21 \pm 1.26$ & $187.59 \pm 0.39$ & $210.90 \pm 0.48$ & 0.032 & 0.0086 & 0.3793\\
        1400 & $185.69 \pm 0.49$ & $192.75 \pm 0.20$ &$204.97 \pm 0.72$ & 0.016 & 0.0084 & 0.1966 \\
    \end{tabular}
    \end{ruledtabular}
    \caption{Sample of 0.2\% offset, 0.5\% EUL, and Johnson-Cook (JC) yield strengths (YS) and errors for Test 1 according to pixel z-level. Note that $\epsilon_{\theta}$ (hoop strain) and $\sigma_{\theta}$ (hoop stress) errors are reported for the data point corresponding to the 0.2\% offset yield point.}
    \label{tab:uncertainty1}
\end{table}

\begin{table}[ht!]
    \centering
    \begin{ruledtabular}
    \begin{tabular}{lllllll}
        \textbf{Z-level} & \textbf{0.2\% YS} & \textbf{0.5\% YS} & \textbf{JC YS} & \textbf{Pixel error} & \textbf{$\epsilon_{\theta}$ error} & \textbf{$\sigma_{\theta}$ error} \\
         & \textbf{[MPa]} & \textbf{[MPa]} & \textbf{[MPa]} & \textbf{[cm]} &  & \textbf{[MPa]} \\
        850 & $190.16 \pm 1.82$ & $190.79 \pm 0.33$ & $216.94 \pm 0.01$ & 0.023 & 0.0085 & 0.3283 \\
        1100 & $183.89 \pm 9.68$ & $186.26 \pm 0.27$ & $217.23 \pm 0.01$ & 0.022 & 0.0084 & 0.2640 \\
        1400 & $174.69 \pm 1.15$ & $70.01 \pm 0.12$ & $20.00 \pm 18.59$ & 0.026 & 0.0083 & 0.3021 \\
    \end{tabular}
    \end{ruledtabular}
    \caption{Sample of 0.2\% offset, 0.5\% EUL, and Johnson-Cook (JC) yield strengths (YS) and errors for Test 2 according to pixel z-level. Note that the ``top'' z-level is 850 for Test 2. The $\epsilon_{\theta}$ (hoop strain) and $\sigma_{\theta}$ (hoop stress) errors are reported for the data point corresponding to the 0.2\% offset yield point.}
    \label{tab:uncertainty2}
\end{table}

The hoop stress--strain curves in Fig.~\ref{fig:diameters_result}cd provide physical values which can be compared to the Ansys simulations. Hoop strain $\epsilon_{\theta}$ is an expansion ratio, measuring the change in outer diameter $D(\tau)$ relative to the initial unpressurized external cylinder diameter $D_0 = 152.6$ mm, calculated as

$$\epsilon_{\theta}(\tau) = \frac{D(\tau) - D_0}{D_0}.$$

Hoop stress $\sigma_{\theta}$ measures the expansion parallel to the cylinder's circumference and for a thin-walled cylinder (wall thickness $t = 1.5$ mm $\ll D_0$) at pressure $P$ is described by the following equation:

$$\sigma_{\theta}(\tau) = \frac{PD(\tau)}{2t}.$$

The curves in Fig.~\ref{fig:diameters_result}cd were produced using pixel z-level 1100, roughly corresponding to the middlemost part of the cylinder in both tests and representative of curve behavior. Due to the corruption described in the following subsection, some of the extracted timestamps did not align properly with those provided with the pressure data; to address this, we performed filtering on the raw diameter data so that diameters (i) below a specified cylinder diameter and (ii) changing by more than a set threshold from the previous recorded diameter would not contribute to the final curve. Hoop strain and stress were then calculated for the pressure data and filtered diameter data using the above equations.

The yield point, a critical input parameter for validating the Ansys RT mechanical simulations, indicates the point at which the deformation transitions from being elastic (temporary) to plastic (permanent); yield strength (YS) is the corresponding stress at this point. We used three independent methods to calculate yield strength: 0.2\% offset, 0.5\% extension under load (EUL), and the Johnson-Cook (JC) Plasticity Model~\cite{Stopka, Voyiadjis2005}. The 0.2\% offset YS was calculated by fitting a line to the elastic region of the stress--strain curve (approximated as being up to 180 MPa, based on various stress--strain curves), shifting the intercept to align with strain of $0.002$, and taking the YS for the method to be the intersection of the shifted line and the data. 0.5\% EUL YS was calculated as the stress at strain of $0.005$. JC YS was calculated by fitting the JC model to the data in the plastic region (approximated as being above 180 MPa) and extracting the value of the $A$ fit parameter. For 0.2\% and JC YS values, both strain and stress errors were propagated through the fits. A sample of YS values for all three methods are reported in Table~\ref{tab:uncertainty1} and Table~\ref{tab:uncertainty2} for Tests 1 and 2, respectively. While these values are largely much higher than the expected value for the material ($\sim68$~MPa)---with the significant exception of the 0.5\% EUL and Johnson-Cook methods for z-level 1400 in Test 2, whose results we suspect may have been impacted by obstructions during segmentation---the specific geometry and cold working can raise the yield strength of copper. This discrepancy could further be accounted for in the work hardening of the copper during the expansion phase of the hydrostatic testing. 

\begin{figure}[htb!]
    \centering
    \includegraphics[width=\textwidth]{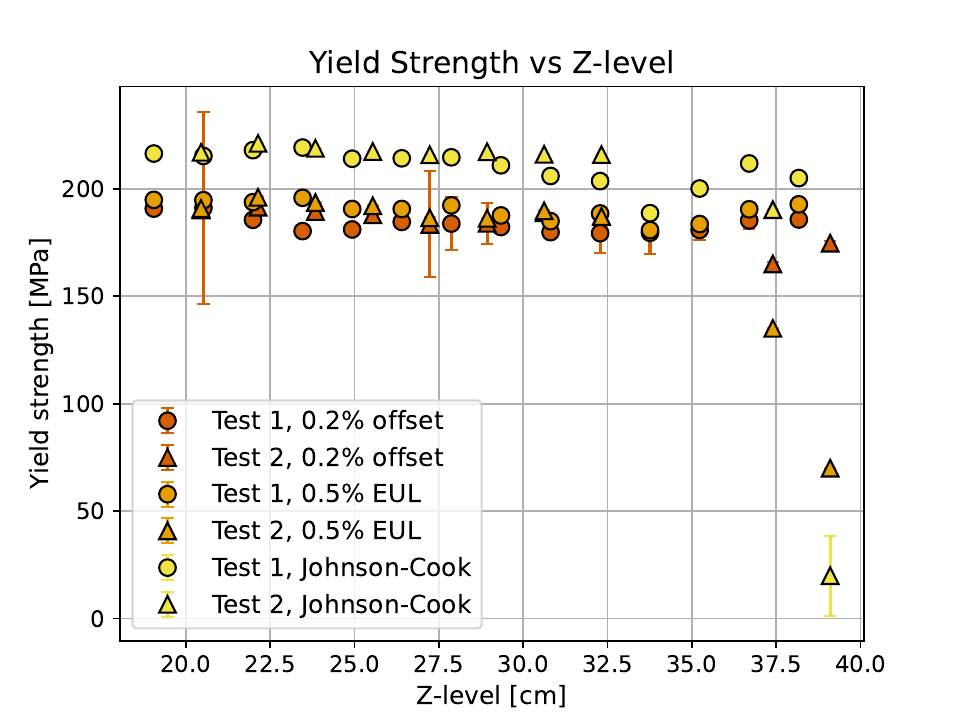}
    \caption{Yield strength extracted from the stress--strain curves at each z-level for both Tests 1 and 2. Yield strength calculation methods are indicated by color: 0.2\% offset (red), 0.5\% EUL (orange), JC (yellow); test is indicated by marker shape: Test 1 (circle), Test 2 (triangle). The z-level is measured as the depth from the top of the cylinder. The gap in Test 2 data (from roughly 32.5 cm to 37 cm) arises from the z-level choices in Fig.~\ref{fig:test1outlines}d, themselves impacted by obstructions (tape, wires) in the video.}
    \label{fig:yieldstrength_zlevel}
\end{figure}

Using the techniques outlined above we were then able to calculate the yield strength of the material at each z-level of the cylinders for both Tests 1 and 2 as shown in Fig.~\ref{fig:yieldstrength_zlevel}. Most clearly in Test 2 the yield strength falls off closer to the bottom endcap, but shows a consistent yield strength for the OFHC copper in the upper and middle sections, which became the most extended during testing. The 0.2\% offset and 0.5\% EUL methods generally fall within 180--195 MPa across both tests.

We calculated the uncertainty in diameter extraction (pixel error) by measuring the standard deviation of diameters for a period before the start of pressurization---for Test 1 this region was 13:59:58 to 14:05:30, while for Test 2 we used 13:39:06 to 13:42:38. Specific to Test 1 was that the only available footage began after the start of pressurization (13:45:40). As such, the Test 1 region was chosen to align with the start of the footage, where although the cylinder was actively being pressurized there was no visible deformation in the copper. Measurement (pixel) errors and propagated stress/strain errors for each test according to z-level are also presented in Tables~\ref{tab:uncertainty1} and ~\ref{tab:uncertainty2}.

\subsection{Corrupted Frames}

\begin{figure}[h]
    \centering
    \begin{minipage}[b]{0.48\textwidth}
        \centering
        \includegraphics[width=\textwidth]{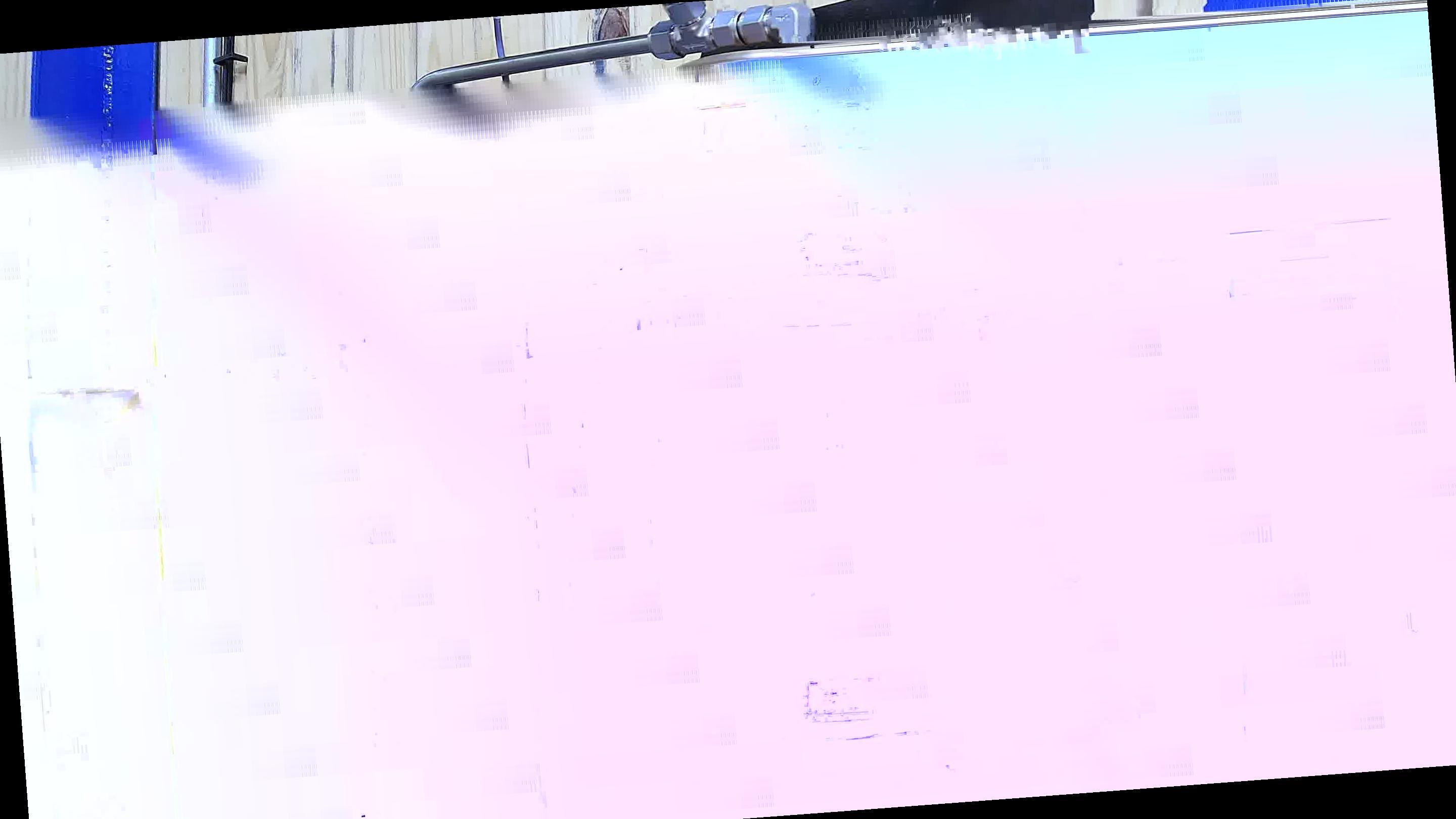}
        \\ (a)
    \end{minipage}
    \hfill
    \begin{minipage}[b]{0.48\textwidth}
        \centering
        \includegraphics[width=\textwidth]{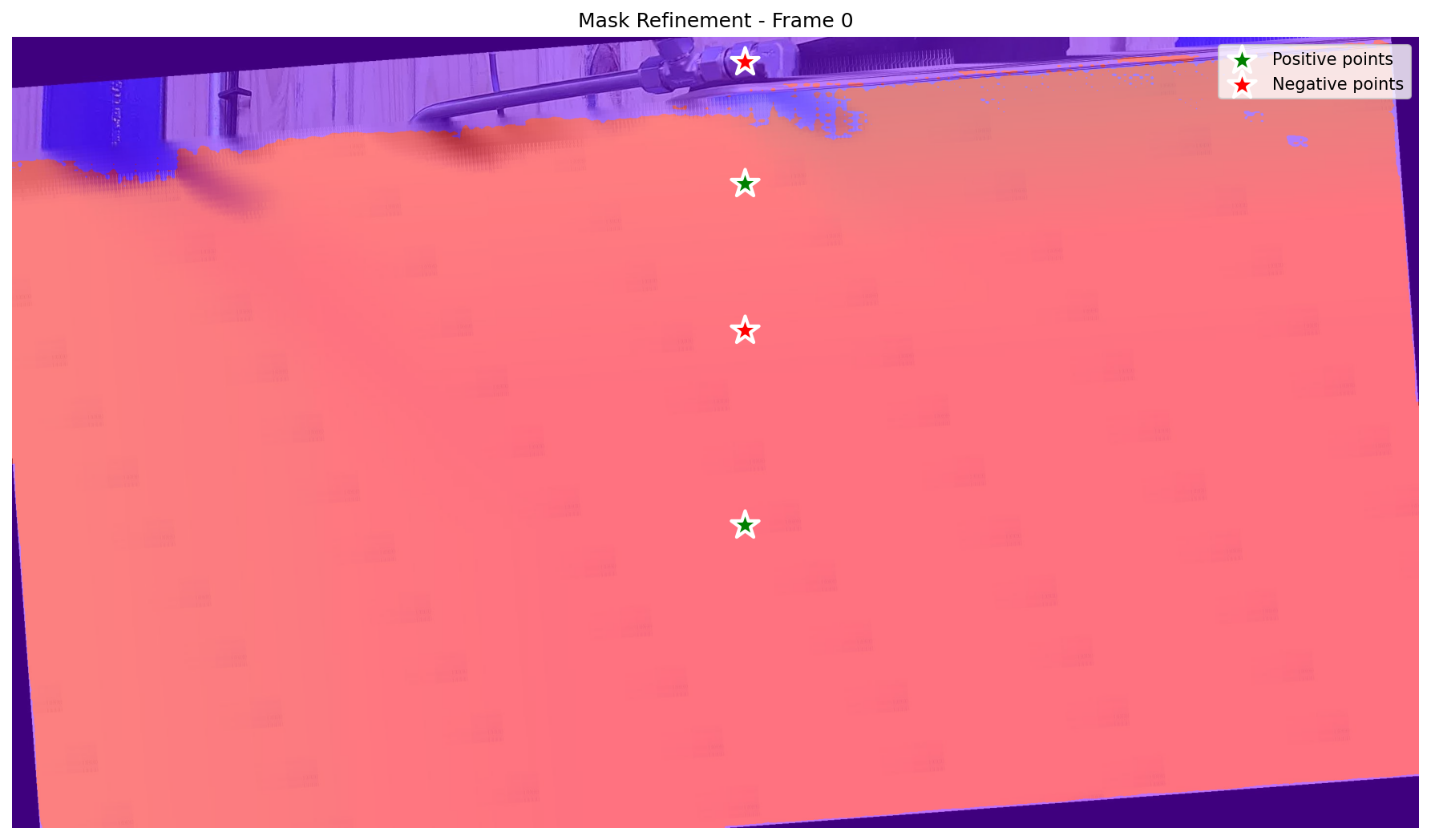}
        \\ (b)
    \end{minipage}
    \caption{(a) Raw Test 1 image from the corrupted region for timestamps 14:22:37 
to 14:22:42. (b) Test 1 image from the corrupted region for timestamps 14:22:37 
to 14:22:42 with segmentation mask applied. Stars indicate fixed points in 
the image that were to be included/excluded from any segmentation mask.}
    \label{fig:corruption_result}
\end{figure}

During the analysis of the Test 1 dataset, we observed instability in the diameter measurements prior to the onset of physical deformation around 14:18. Subsequent investigation attributed these fluctuations to intermittent frame corruption within the Test 1 source video. A representative example of such corruption is presented in Fig.~\ref{fig:corruption_result}. During these corruption events, visual information regarding the cylinder's geometry was effectively lost, preventing accurate diameter extraction. We performed a manual inspection of the footage to identify locations of human-visible corrupted frames, which are masked red in Fig.~\ref{fig:diameters_result}a. The results imply some temporal correlation between the observed diameter instabilities and the occurrence of frame corruption. Conversely, as shown in Fig.~\ref{fig:diameters_result}b, the Test 2 dataset exhibited higher video quality and stability and was largely albeit not wholly unaffected by these artifacts.

To account for occasions when the SAM2 segmentation fails and timestamps are unrecoverable, the agent adds a placeholder value to maintain chronological and sequential order. We also observed that corrupted frames critically impacted subsequent diameter extraction; specifically, a segmentation attempt on a single corrupted frame as in Fig.~\ref{fig:corruption_result}b propagated errors throughout the entire chunk. While large deviations in diameters flagged potential issues, identifying frozen frames required manual verification of timestamp discontinuities. Therefore, these corrupted intervals were excluded from the final analysis to ensure consistency. Our future work will involve building additional tools to cross-check and remove corrupted frames.

\subsection{Cross-Validation with Non-Agentic Methods}
As a cross-validation method, we also developed a non-agentic pipeline for the same diameter- and timestamp-extraction tasks. While this manual pipeline served as a successful sanity check, it necessitated significant human attention and fine-tuning: as there was no memory management in the non-agentic pipeline, frames were manually chunked and then re-chunked upon OOM error. The pipeline was completely unautomated, with segmentation masks applied manually, and the kernel had to be restarted after diameter extraction for each chunk so as to clear memory. Furthermore, timestamp extraction relied solely on EasyOCR, resulting in a number of poorly-extracted timestamps (particularly for Test 2)---and timestamp validation was performed entirely by hand.

\begin{figure}[ht]
    \centering
    \begin{minipage}[b]{0.48\textwidth}
        \centering
        \includegraphics[width=\textwidth]{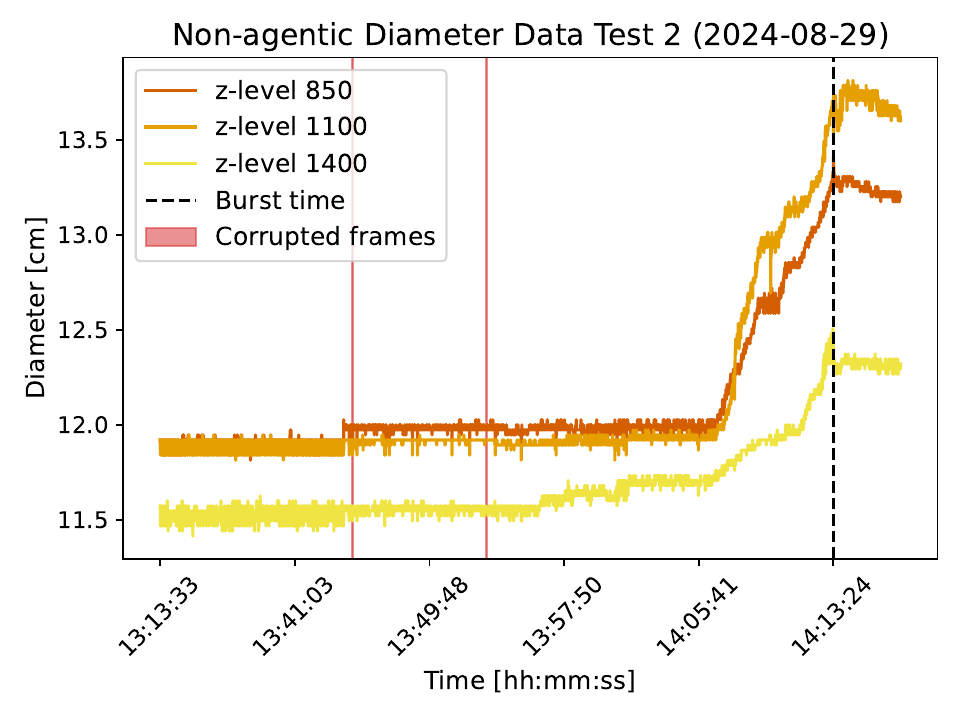}
        \\ (a)
    \end{minipage}
    \hfill
    \begin{minipage}[b]{0.48\textwidth}
        \centering
        \includegraphics[width=\textwidth]{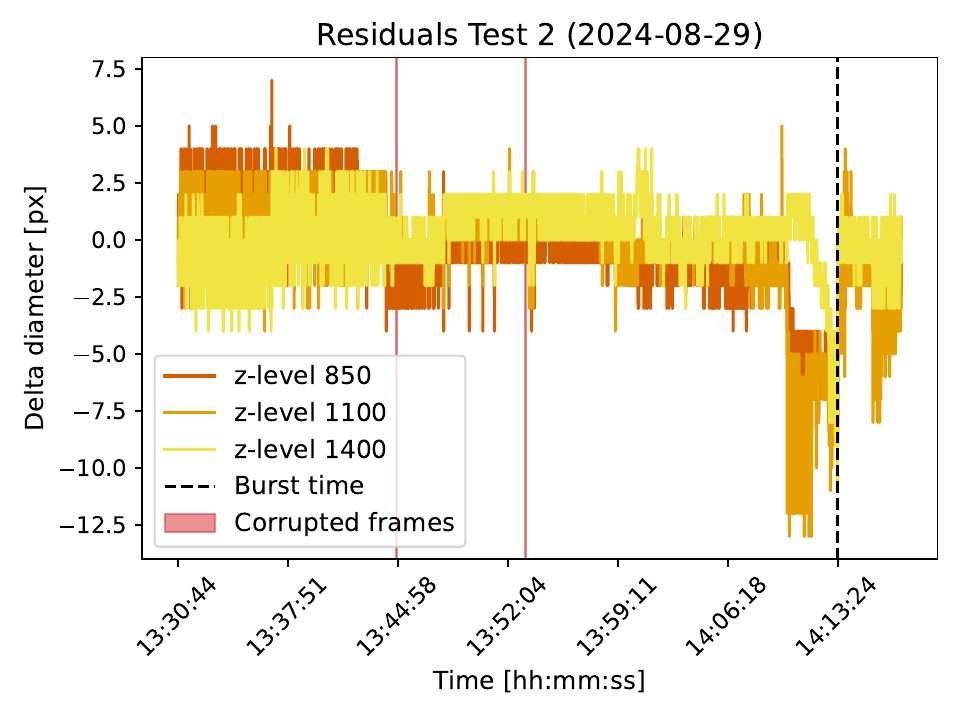}
        \\ (b)
    \end{minipage}
    \caption{(a) Time evolution of LoFi cylinder diameters extracted by the 
non-agentic pipeline for Test 2. (b) Raw residuals comparing the agentic 
and non-agentic pipelines.}
    \label{fig:comparison_result}
\end{figure}

The results yielded by this non-agentic approach for Test 2 are presented in Fig.~\ref{fig:comparison_result}a, while Fig.~\ref{fig:comparison_result}b displays a residual plot comparing these measurements against the agentic pipeline. A good agreement is reached between the two versions, with the absolute value of the residual mostly within $\pm 5$ pixels, except near the burst time where we have an extremely dynamic image environment. We suspect that the majority of deviations across methods result largely from differences in chunk size, as two different frames can produce two different segmentation masks which are then propagated across the rest of the chunk and influence the positions of the LoFi cylinder edges.
\subsection{Comparison to LEGEND Ansys Simulations}

\begin{figure}[htbp]
    \centering
    \begin{minipage}[b]{0.48\textwidth}
        \centering
        \includegraphics[width=\textwidth]{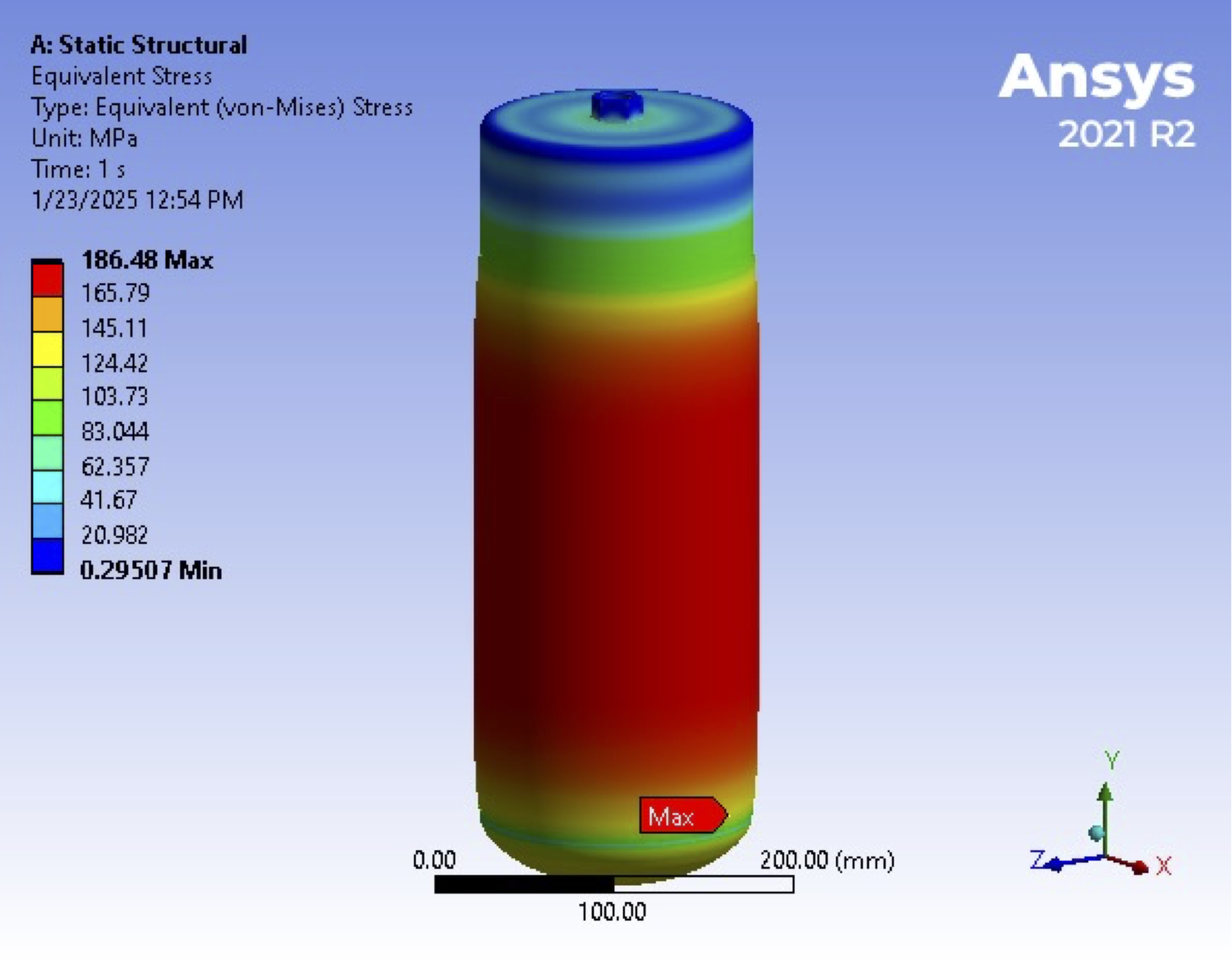}
        \\ (a)
    \end{minipage}
    \hfill
    \begin{minipage}[b]{0.48\textwidth}
        \centering
        \includegraphics[width=\textwidth]{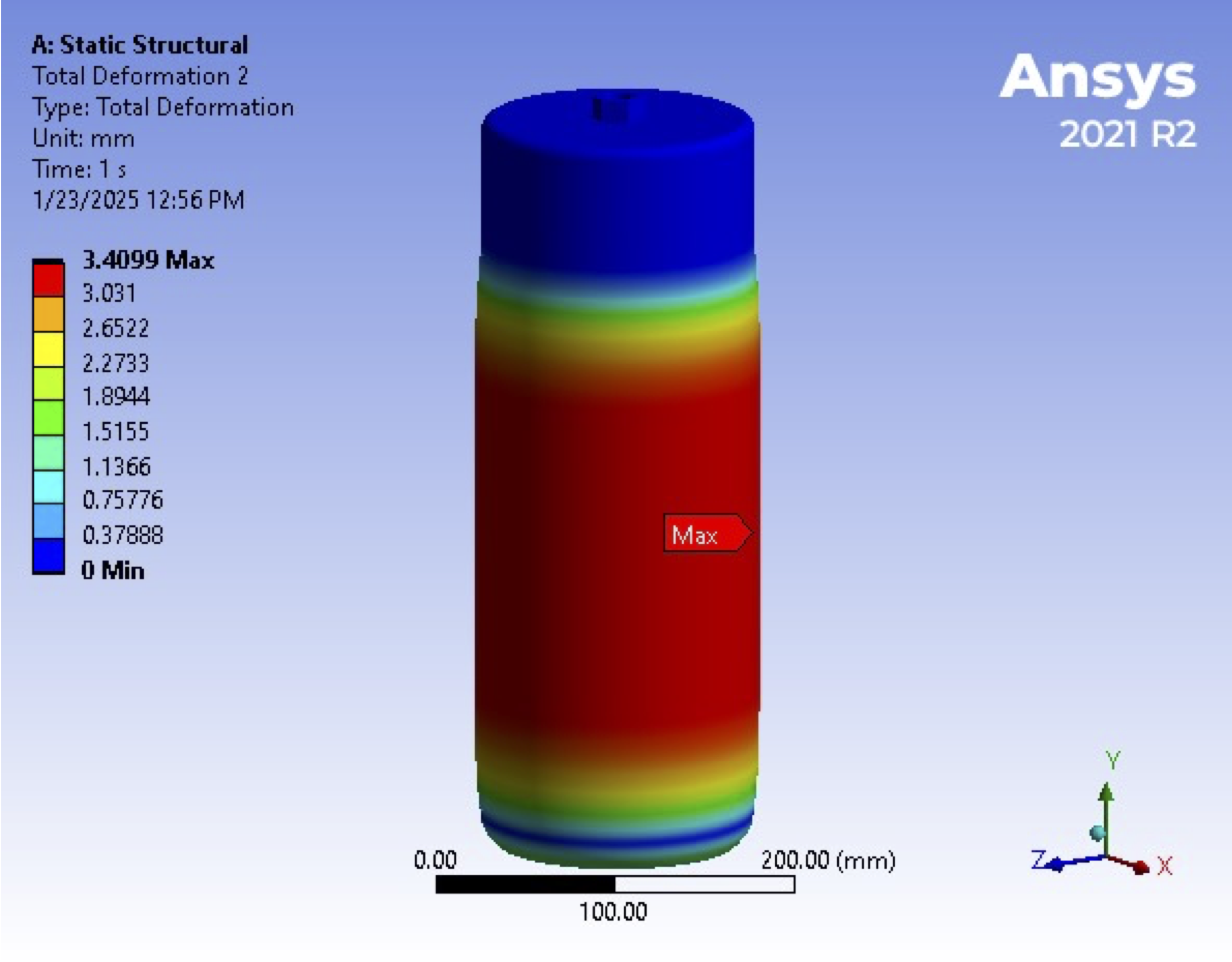}
        \\ (b)
    \end{minipage}
    \caption{Results of the total (a) structural stress and (b) deflection of 
the LoFi cylinder when put under the same conditions as Tests 1/2 (pressurized 
up to 4.14 MPa). Performed in Ansys~\cite{ansysmech2024}.}
    \label{fig:ansyscomparision}
\end{figure}
\vspace{-0.5em}

As part of the LEGEND RT campaign, we undertook a series of mechanical simulations, performed in Ansys, to verify the viability of the overall RT design, with the mechanical properties for the copper in particular taken from the literature. These LoFi tests provided an opportunity to cross-check the material properties of copper with the experimental data collected during the hydrostatic testing. The raw results of these simulations are shown in Fig.~\ref{fig:ansyscomparision}, which showcases some of the differences that have had to be accounted for in subsequent mechanical simulations of the RT design. 

While the maximum overall stress level did closely match the values acquired from the Test 1 and 2 videos, the maximum deformation in the simulation (3.41 mm) failed to align with the final maximum deformation observed in our tests ($\sim45$ mm). This can be attributed to the fact that our initial Ansys simulations were designed to simulate linear, steady-state effects at equilibrium. During the final $\sim5$ minutes of Tests 1/2 (see Fig.~\ref{fig:diameters_result}ab), it is clearly noticeable that the diameters started rapidly changing, indicating that the cylinders had entered the dynamic plastic and eventually the catastrophic burst stage. In the earlier part of the testing there was better agreement as the cylinder was elastic and the total deformation remained below 5 mm.

As previously discussed, we attribute our higher, calculated yield strengths to work hardening of the copper which was cold-rolled during the manufacturing process to its final thickness of 1.5 mm. Current efforts of the RT team are focusing on incorporating these updated material values for the yield of OFHC copper into the final RT design, as well as exploring the possibility of performing similar testing at cryogenic temperatures to better mimic the final expected conditions of the RT during deployment. 


\section{Future Improvements}\label{sec:improvements}
As this was the first application of this technique for extracting
yield strength from a hydrostatic pressure test, we note several areas of improvement for subsequent measurement runs. The principal limitation came from the source video itself, which afforded only a single, resolution-limited view of the LoFi cylinder. Future measurements would therefore benefit from multiple camera angles, in particular a view from above to capture any non-axisymmetric deformations arising during the test. Higher pixel resolution would also in turn sharpen spatial resolution, reducing the uncertainty on each stress--strain point, and higher-quality filming equipment would reduce if not eliminate footage corruption. Dimensional accuracy, and by extension the yield strength parameter extraction, could be improved further by placing known dimensional standards within the frame of each video, giving a consistent benchmark for the agent's measurements across the entire field of view. Finally, the yield strengths calculated from the diameters should be cross-checked against independent measurements from strain gauges placed at specific height as a validation step.

\section{Conclusion and Outlook}\label{sec:conclusion}
In order to meet its physics goals, the LEGEND-1000 experiment will build a 6-m-tall, 2-m-diameter tube with 1.5-mm-thick walls partly out of high-purity copper that meets strict background requirements, pushing the limits of current manufacturing techniques. As part of the larger LEGEND-1000 R\&D hardware characterization campaign, a series of copper test cylinders were commissioned and hydrostatically pressure-tested until the weld seams failed; unexpectedly, the strain gauges attached to the cylinders also failed, returning poor data if not detaching themselves entirely. In order to then obtain material strain parameters, we developed a computer vision-based AI agent to autonomously determine the cylinder diameters from videos of the tests, which were synchronized with the pressurization data. The agent was tasked with extracting diameters and timestamps from each video while having to overcome various anomalies found in the data, such as slanted/corrupted frames or incomplete timestamp extraction. By adopting an agentic workflow, we significantly reduced the reliance on human intervention and monitoring in favor of automating the process for dealing with these irregularities. We have shown that our agentic framework can identify the test cylinder in our testing videos and measure the diameters of the cylinder as well as the timestamps in any given still. The agentic framework performance was further validated through a comprehensive analysis of the agent's behavior and uncertainty quantification, benchmarking it directly against our non-automated pipeline and showing an agreement at the level of $\pm$ a few pixels. Finally, we were able to calculate consistent yield strengths from both of the testing videos, which will contribute to the overall engineering designs for the LEGEND-1000 experiment.



A key advantage of our vision AI agent is its high degree of generalizability; it establishes a robust framework applicable to any experimental task requiring precise distance or displacement measurements from video data, expanding on traditional techniques relying on strain gauges which in this instance would only have been able to provide limited, highly-localized data. Building on this, we plan to extend the use of this agent to a broader class of applications in large-scale physics experiments including real-time detector monitoring, mechanical structure assessment, and material layer thickness extraction. This work demonstrates the power of AI agents in experimental particle physics and their applications beyond simply physics analysis tools.

\begin{acknowledgments}
We wish to thank Matthew Busch and Bernhard Schwingenheuer for their technical assistance and industry connections to aid with the design and manufacture of the LoFi test cylinders. We gratefully acknowledge the support of the U.S. Department of Energy through the Los Alamos National Laboratory Laboratory Directed Research and Development Program. This work was supported by resources of the National Energy Research Scientific Computing Center, which is supported by the Office of Science of the U.S. Department of Energy under Contract No. DE-AC02-05CH11231. The work at UCSD is supported by DE-AC02-05CH11231 Subaward 7836985 and Aobo Li's startup. 
\end{acknowledgments}

\bibliography{biblist}

\appendix
 
\clearpage
\section{AI Glossary}\label{app:glossary}
We present a short glossary of AI-related terms mentioned in this work. Terms are listed hierarchically relative to one another, not in order of appearance.
\begin{itemize}
    \item \textbf{Foundation model}: a large-scale AI system trained on various unlabeled data (e.g., text, images, audio) to serve as a basis for more-specialized applications.
    \item \textbf{Large Language Model (LLM)}: a kind of foundation model specifically trained on large bodies of text to specialize in understanding, generating, and translating natural language. Popular examples include ChatGPT, Gemini, and Claude.
    \item \textbf{AI agent}: a form of high-level automation, with a large language model as the central reasoning engine. The LLM interfaces with tools according to a system prompt to complete desired tasks.
    \item \textbf{Computer Vision (CV)}: a broad field of AI focused on helping computers ``see,'' enabling them to perform tasks such as interpreting and understanding visual input like images and videos.
    \item \textbf{Optical Character Recognition (OCR)}: a specific type of computer vision focused on text extraction from non-typed text sources (e.g., images, PDFs, handwritten documents).
\end{itemize}

\clearpage
\section{Tools}\label{app:tool_list}

Central to agent setup is the definition of tool functionality via natural language prompts. For instance, the instruction for the \texttt{plan\_adaptive\_chunk\_processing} tool is defined as follows:

\begin{verbatim}
"""
Plan adaptive chunk processing that adjusts chunk size based on memory usage.
Starts with initial chunk size, then adapts based on actual memory consumption.

Args:

total_frames: Total number of frames in the video
starting_chunk_size: Initial chunk size to start with (default: 250)
source_dir: Directory with all frames

Returns:
Dictionary with initial plan
"""
\end{verbatim}
 
Tables \ref{tab:video}, \ref{tab:diameter}, and \ref{tab:timestamps} list the tools used within each subagent, alongside a description of their usage.

\begin{table}[ht!]
    \centering
    \begin{ruledtabular}
    \begin{tabular}{ll}
        \textbf{Tool} & \textbf{Usage} \\
        \hline
        \texttt{decompose\_video\_to\_frames} & \parbox[t]{0.55\textwidth}{Decompose a video into individual frames using FFmpeg.} \\[6pt]
        \texttt{calculate\_rotation\_angle} & \parbox[t]{0.55\textwidth}{Calculate rotation angle needed to make the container vertical.} \\[6pt]
        \texttt{rotate\_all\_frames} & \parbox[t]{0.55\textwidth}{Rotate all frames in a directory and save to output directory.} \\
    \end{tabular}
    \end{ruledtabular}
    \caption{Tools and their usages for the Video Agent.}
    \label{tab:video}
\end{table}

\begin{table}[ht!]
    \centering
    {\footnotesize
    \begin{ruledtabular}
    \begin{tabular}{ll}
        \textbf{Tool} & \textbf{Usage} \\
        \hline
        \texttt{check\_gpu\_memory\_status} & \parbox[t]{0.65\textwidth}{Check current GPU memory usage.} \\[6pt]
        \texttt{clear\_gpu\_cache} & \parbox[t]{0.65\textwidth}{Clear GPU memory cache.} \\[6pt]
        \texttt{count\_frames\_in\_directory} & \parbox[t]{0.65\textwidth}{Count frames in a directory.} \\[6pt]
        \texttt{process\_chunk\_with\_sam2} & \parbox[t]{0.65\textwidth}{Process a chunk with SAM2. Uses sequential point application by default for better masks.} \\[6pt]
        \texttt{propagate\_masks\_in\_chunk} & \parbox[t]{0.65\textwidth}{Propagate masks and store results in global context for diameter extraction. Optionally visualizes the mask before propagating.} \\[3pt]
        \texttt{calculate\_diameters\_at\_rows} & \parbox[t]{0.65\textwidth}{Calculate diameters at specified rows using non-agentic pipeline logic.} \\[6pt]
        \texttt{calculate\_pix\_value} & \parbox[t]{0.65\textwidth}{Calculate vertical pixel offset (pix value)---distance from frame top to mask top.} \\[6pt]
        \texttt{plan\_adaptive\_chunk\_processing} & \parbox[t]{0.65\textwidth}{Plan adaptive chunk processing that adjusts chunk size based on memory usage. Starts with initial chunk size, then adapts based on actual memory consumption.} \\[6pt]
        \texttt{validate\_chunk\_boundaries} & \parbox[t]{0.65\textwidth}{Validate that chunk boundaries don't split the protected/corrupted region. The protected region must either be completely outside the chunk or completely inside the chunk.} \\[6pt]
        \texttt{get\_next\_chunk\_adaptive} & \parbox[t]{0.65\textwidth}{Get the boundaries for the next chunk using adaptive sizing. Uses memory usage from previous chunk to determine optimal size.} \\[6pt]
        \texttt{update\_adaptive\_chunk\_size} & \parbox[t]{0.65\textwidth}{Update chunk size for next iteration based on memory usage.} \\[6pt]
        \texttt{append\_measurements\_to\_file} & \parbox[t]{0.65\textwidth}{Append current chunk's measurements to the master pickle file. If file does not exist, creates it. If exists, appends data.} \\[6pt]
        \texttt{cleanup\_chunk\_data} & \parbox[t]{0.65\textwidth}{Clean up chunk processing data after measurements are saved. Deletes inference states, video segments, GPU cache, chunk directories, and recreates SAM2 predictor to prevent memory leaks.} \\[6pt]
    \end{tabular}
    \end{ruledtabular}
    }
    \caption{Tools and their usages for the Diameter Agent.}
    \label{tab:diameter}
\end{table}

\begin{table}[ht!]
    \centering
    \begin{ruledtabular}
    \begin{tabular}{ll}
        \textbf{Tool} & \textbf{Usage} \\
        \hline
        \texttt{count\_frames\_in\_directory} & \parbox[t]{0.55\textwidth}{Count frames in a directory.} \\[6pt]
        \texttt{save\_measurements\_to\_pickle} & \parbox[t]{0.55\textwidth}{Save the populated measurement data to a pickle file.} \\[6pt]
        \texttt{load\_existing\_measurements} & \parbox[t]{0.55\textwidth}{Load existing measurement data from pickle file into context.} \\[6pt]
        \texttt{initialize\_context} & \parbox[t]{0.55\textwidth}{Initialize the context for a new video analysis session.} \\[6pt]
        \texttt{set\_frame\_source} & \parbox[t]{0.55\textwidth}{Set the source directory and filename pattern for reading video frames.} \\[6pt]
        \texttt{set\_timestamp\_region} & \parbox[t]{0.55\textwidth}{Set the region in the frame where timestamps appear for OCR extraction.} \\[6pt]
        \texttt{extract\_timestamps\_with\_easyocr} & \parbox[t]{0.55\textwidth}{Extract timestamps from video frames using EasyOCR with validation.} \\[6pt]
        \texttt{validate\_timestamp\_with\_llm} & \parbox[t]{0.55\textwidth}{Use Claude Haiku's vision capabilities to extract timestamp from a problematic frame.} \\[6pt]
        \texttt{batch\_validate\_timestamps\_with\_llm} & \parbox[t]{0.55\textwidth}{Validate multiple problematic frames in batches using Claude Haiku's vision.} \\[6pt]
        \texttt{populate\_timestamps\_to\_pickle} & \parbox[t]{0.55\textwidth}{Add extracted timestamps to the pickle data structure.} \\[6pt]
    \end{tabular}
    \end{ruledtabular}
    \caption{Tools and their usages for the Timestamp Agent.}
    \label{tab:timestamps}
\end{table}

\clearpage

\end{document}